\documentclass[letter,scriptaddress,twocolumn, pre,showkeys]{revtex4}

    \usepackage{amsmath}
    \usepackage{makeidx}
    \usepackage{amsfonts}
    \usepackage[ansinew]{inputenc}
    \usepackage[usenames,dvipsnames]{pstricks}
    \usepackage{subfigure}
    \usepackage{epsfig}
    \usepackage{pst-grad} 
    \usepackage{pst-plot} 
    \usepackage[colorlinks,hyperindex]{hyperref}
    \hypersetup
    {
        colorlinks,%
        citecolor=black,%
        linkcolor=black,%
        urlcolor=black,%
    }

    \newtheorem{axiom}{Axiom}

\makeindex

\begin{document}

\title{The introduction of symmetry constraints within MaxEnt Jaynes's methodology}

\author{F. Holik}
\email{holik@fisica.unlp.edu.ar}
\affiliation{Center Leo Apostel for Interdisciplinary Studies and,
Department of Mathematics, Brussels Free University
Krijgskundestraat 33, 1160 Brussels, Belgium\\ National University
La Plata \& CONICET IFLP-CCT, C.C. 727 - 1900 La Plata, Argentina}

\author{C. Massri}
\email{cmassri@dm.uba.ar}
\affiliation{Department of Mathematics, University of Buenos Aires \& CONICET IMAS.}

\author{A. Plastino}
\email{angeloplastino@gmail.com}
\affiliation{National University La Plata \& CONICET IFLP-CCT, C.C.
727 - 1900 La Plata, Argentina}

\date{today}

\begin{abstract}
We provide a generalization of the approach to geometric probability
advanced by the great mathematician Gian Carlo Rota, in order to
apply it to generalized probabilistic physical theories. In
particular, we use this generalization to provide an improvement of
the Jaynes' MaxEnt method. The improvement consists in providing a
framework for the introduction of symmetry constrains. This allows
us to include group theory within MaxEnt. Some examples are
provided.
\end{abstract}


\maketitle

\section{Introduction}

\noindent Jaynes' MaxEnt approach is a statistical approach in which
probability notions become of the essence
\cite{Jaynes-1957a,Jaynes-1957b,Presse-MaxEntMaxCalRevModPhys}.
Thus, new viewpoints regarding probability are susceptible of
modifying the MaxEnt approach. We center our present efforts on the
notion of geometric probability, characterized by Gian Carlo Rota as
the study of invariant measures \cite{Rota1998,RotaBook}. This idea
has lead to interesting mathematical problems, which have defined a
rich field of study. {\it In this work, we provide a generalization
of the Rota's axioms in order to find a physical characterization of
the problem of looking for generalized probabilities in the spirit
of Jaynes's MaxEnt approach}. As it is well known, this technique
relies in the determination of the less unbiased distribution
compatible with the known data, by appealing to the maximization of
the entropy \cite{Jaynes-1957a,Jaynes-1957b} and has manifold
applications in diverse fields of research
\cite{Cavagna-2014-BirdsMaxEnt,BerettaMaxEnt-2014,Sinatra-MaxEntRandomWalks-2011,Dirks-MonteCarloMaxEnt,Trovato-Reggiani-MaxEnt-PRE-IdenticalParticles,Diambra-Plastino-PRE-MaxEnt-1995,Rebollo-Neira-Plastino-PRE-MaxEnt-2001,Diambra-Plastino-PRE-MaxEnt-1996,Goswami-Prasad-PRD-MaxEnt-2013,Cofre-Cessac-MaexEntNeuralNetworks,Lanata-MaximumEntanglement-2014,Trovato-MaxEntFractionalExclusion,Goncalves-2013-MaxEntTomography,Canosa-Plastino-Rossignoli-PRA-MaxEnt,Plastino-Portesi-PRA-MaxEnt,Tkacic-MaxEntNeuralNetwork-2013,Knuth:2014Zitter}
(see \cite{Presse-MaxEntMaxCalRevModPhys} for a complete review).
Our methodology can be used to find a derivation of both classical
and quantum statistical mechanics as well.

{\it Our treatment reformulates the MaxEnt approach in geometric
probability terms, allowing for the inclusion of group actions
representing physical symmetries}. In this framework, states of a
physical system are regarded as \emph{invariant measures} over
general orthomodular lattices (a lattice is a partially ordered set
with unique least upper bounds and greatest lower bounds. For
details see, for instance,
\cite{KalmbachOrthomodularLatticesBook,Holik-Zuberman-2013}). The
determination of invariant measures under the action of groups
representing physical symmetries is of interest in many research
fields, as for example, in the problem of the determination of
equilibrium states in equilibrium statistical mechanics
\cite{Landford-Robinson-JMP1968,Lanford1968,lanford1969}. We also
provide an improvement on the treatment of constrains by formulating
the problem in the rigorous basis of measure theory, and allowing
for them a more general character than mere mean values. We show as
well that the introduction of group actions reduces the
dimensionality of the mathematical variety on which the maximization
process takes place. This economizes computational resources. We
demonstrate that this economization can be estimated for certain
examples. Finally, we provide some examples and specify conditions
under which solutions for our method exist.

The paper is organized as follows. In
Section~\ref{s:GeometricProbability}, we introduce the elementary
notions of geometric probability theory
following\cite{Rota1998}\footnote{The reader familiarized with this
theory can skip this Section.}. In Section \ref{s:EventStructures},
we review event structures appearing in both in quantum and
classical mechanics ---and their associated probabilities---, and in
more general probabilistic settings as well. In Section
\ref{s:NewSetOfAxioms}, we propose a generalization of geometric
probability theory which allows one to describe physical systems. In
Section \ref{s:CovarianceAndSymmetries}, we explain how covariance
conditions and physical symmetries can be accommodated by our
 conceptual framework.  In Sections \ref{s:CoherentStates} and
\ref{s:BellInequalities}, we show how to describe, in our framework,
quantum coherent states and the correlations appearing in the
no-signal polytope. Finally, we draw some conclusions in Section
\ref{s:Conclusions}.

\section{Geometric Probability}\label{s:GeometricProbability}

\noindent In his classical approach to geometric probability
\cite{Rota1998,RotaBook}, Gian Carlo Rota introduces the problem
of invariant measures as follows. First, one looks for a measure
$\mu:\Sigma\longrightarrow \mathbb{R}_{\geq 0}$, defined on a
sigma algebra $\Sigma\subseteq\mathcal{P}(\mathbb{R}^{n})$,
satisfying the following axioms

\begin{axiom}[R1]\label{a:RotaAxiom1}
$$\mu(\emptyset)=0$$
\end{axiom}

\noindent where $\emptyset$ denotes the empty set. If $A$ and $B$
are measurable sets:

\begin{axiom}[R2]\label{a:RotaAxiom2}
$$\mu(A\cup B)=\mu(A)+\mu(B)-\mu(A\cap B)$$
\end{axiom}

\noindent For Boolean algebras, the above axiom is equivalent to the
sum rule

\begin{equation}\label{a:SumRule}
\mu(A\cup B)=\mu(A)+\mu(B)
\end{equation}

\noindent for $A$ and $B$ disjoint. The following axiom has to do
with the invariance of measures (therefore, the name \emph{invariant
measures}):

\begin{axiom}[R3]\label{a:RotaAxiom3}
The measure of a set $A$ does not depend on the position of $A$; in
other words, if $A$ can be rigidly transformed into $B$, then, $B$
and $A$ have the same measure.
\end{axiom}

\noindent Notice that the last axiom involves the action of a group,
namely, the Euclidean group $\mathrm{E}_{0}$ of rotations and
translations in Euclidean space. The last axiom specifies a
normalization for a given measure; we must pick a special subset and
establish its measure. Let us choose the set of parallelotopes $P$
with orthogonal side lengths $x_{1},\ldots,x_{n}$ and impose the
constrain:

\begin{axiom}[R4]\label{a:RotaAxiom4}
$$\mu(P)=x_{1}x_{2}\cdots x_{n}$$
\end{axiom}

\noindent The above axioms yield the usual Lebesgue measure on
$\mathbb{R}^{n}$. Rota poses the question of what happens if the
normalization \textbf{Axiom} \ref{a:RotaAxiom4} is changed. Instead
of \textbf{Axiom} \ref{a:RotaAxiom4}, one could use one of the
following polynomials

\begin{subequations}
\begin{equation}\label{e:Norme1}
e_{1}(x_{1}, x_{2},\ldots,x_{n}) = x_{1} + x_{2} + \ldots + x_{n}
\end{equation}
\begin{equation}\label{e:Norme2}
e_{2}(x_{1},x_{2},\ldots,x_{n}) = x_{1}x_{2} + x_{1}x_{3} + \ldots +
x_{n-1}x_{n}
\end{equation}
\centering\vdots
\begin{eqnarray}\label{e:NormeMedium}
e_{n-1}(x_{1}, x_{2},\ldots,x_{n})&=& x_{2}x_{3}\cdots x_{n}+
x_{1}x_{3}x_{4}\cdots x_{n}+ \nonumber\\ &+& \ldots +
x_{1}x_{2}\cdots x_{n-1}
\end{eqnarray}

\begin{equation}
e_{n}(x_{1}, x_{2},\ldots,x_{n}) = x_{1}x_{2}\cdots x_{n}
\end{equation}
\end{subequations}

\noindent Indeed, the symmetric polynomial $e_{n}$ is coincident
with the normalization of \textbf{Axiom} \ref{a:RotaAxiom4}.
\emph{Geometric probability studies the conditions under which these
measures exist, and how they can be used to generate more general
ones \cite{Rota1998,RotaBook}.}

Geometric probability theory can be also used for studying invariant
measures in Grassmannians. A complete introduction to the subject
can be found in \cite{RotaBook}. In the following Section, we will
review the formulation of the axioms of a non-commutative
probability calculus, i.e., probabilities which generalize
Kolmogorov's \cite{KolmogorovFoundationsOfProbability} axioms to
non-Boolean settings
\cite{RedeiSummersQuantumProbability,RedeiQuantumLogicInAlgebraicApproach}.

\section{Event Structures}\label{s:EventStructures}

When faced with a concrete physical problem, we are interested in
determining the probabilities of certain events of interest. An
event will be the definite outcome of a certain experiment for which
we can determine the answer with certainty. As an example, we can
think about the detection of a particle (classical or quantal) in a
certain region of space-time, and the probability for this event to
occur.

It happens that events of a physical system can be endowed with
definite mathematical structures
\cite{DallaChiara-Giuntini-Greechie-Book,BeltramettiCassinelliBook,RedeiQuantumLogicInAlgebraicApproach};
if the particle is classical, events may be represented as
measurable subsets of the phase space $\Gamma$. Measurable subsets
of phase space form a well known structure, namely, a \emph{Boolean
algebra}
\cite{BooleAnInvestigation,DallaChiara-Giuntini-Greechie-Book} that
we will denote by $\mathcal{P}(\Gamma)$\footnote{A Boolean lattice
will be a partially ordered set for which i) the least upper bound
(disjunction) and maximum lower bound (conjunction) exists for every
pair of elements; ii) it is orthocomplemented; iii) it is
distributive. A typical model for a Boolean lattice will be that of
the subsets of a given set, with set intersection as conjunction,
set union as disjunction, and set theoretical complement as
orthocomplementation \cite{DallaChiara-Giuntini-Greechie-Book} (see
also \cite{Knuth-Skilling-Axioms-2012} for a study of the algebraic
symmetries of Boolean lattices).}.

On the other hand, as shown by Birkhoff and von Neumann
\cite{Birkhoff-vonNeumann-LogicOfQm}, events associated to a quantum
particle will be naturally represented by \emph{projection
operators}, specifically those  associated to the spectral
decomposition of self adjoint operators representing physical
observables. Unlike the classical Boolean case, projections of a
Hilbert space form an \emph{orthomodular lattice}
$\mathcal{P}(\mathcal{H})$, which can be shown to be
non-distributive
\cite{DallaChiara-Giuntini-Greechie-Book,BeltramettiCassinelliBook,Birkhoff-vonNeumann-LogicOfQm}
(and thus, not Boolean)\footnote{An orthomodular lattice will be an
orthocomplemented lattice for which a condition weaker than
distributivity holds (see for example
\cite{KalmbachOrthomodularLatticesBook,RedeiSummersQuantumProbability,Holik-Zuberman-2013}).
Boolean algebras are always orthomodular lattices, but the converse
is not true \cite{DallaChiara-Giuntini-Greechie-Book}. For
$\mathcal{P}(\mathcal{H})$, conjunction is given intersection,
disjunction by closure of direct sum, and orthocomplementation by
orthogonal complement of the closed subspaces associated to each
projection operator (projection operators can be put in one to one
correspondence with closed subspaces of $\mathcal{H}$).}. This
important mathematical difference between classical and quantum
theories is the direct consequence of the incompatibility of
complementary observables in QM.

\subsection{Classical case}

To illustrate these ideas, let us start by considering the phase
space $\mathbb{R}^{2n}$ of a classical system. If $f$ represents an
observable quantity, the proposition ``the value of $f$ lies in the
interval $\Delta$", defines an \emph{event} $f_{\Delta}$, which can
be represented as the measurable set $f^{-1}(\Delta)$ (the set of
all states which make the proposition true). If the probabilistic
state of the system is given by $\mu$, the corresponding probability
of occurrence of $f_{\Delta}$ will be given by
$\mu(f^{-1}(\Delta))$. As an example, consider the energy of an
harmonic oscillator. The proposition ``the energy of the oscillator
equals $\varepsilon$" corresponds to an ellipse in phase space for
each possible value of $\varepsilon$.

There is a strict correspondence between a classical probabilistic
state and the axioms of classical probability theory. Indeed, the
axioms of Kolmogorov \cite{KolmogorovFoundationsOfProbability}
define a probability function as a measure $\mu$ on a sigma-algebra
$\Sigma$ such that

\begin{subequations}\label{e:kolmogorovian}
\begin{equation}
\mu:\Sigma\rightarrow[0,1]
\end{equation}
\noindent which satisfies
\begin{equation}
\mu(\emptyset)=0
\end{equation}
\begin{equation}
\mu(A^{c})=1-\mu(A),
\end{equation}

\noindent where $(\ldots)^{c}$ means set-theoretical-complement.
For any pairwise disjoint denumerable family $\{A_{i}\}_{i\in I}$,

\begin{equation}\label{e:sigmaAdditivityKolomogorov}
\mu(\bigcup_{i\in I}A_{i})=\sum_{i}\mu(A_{i}).
\end{equation}

\end{subequations}

\noindent A state of a classical probabilistic theory will be
defined as a Kolmogorovian measure with
$\Sigma=\mathcal{P}(\Gamma)$. The reader will also notice the
analogy between the first two Rota's \textbf{Axioms}
\ref{a:RotaAxiom1} and \ref{a:RotaAxiom2} and the axioms of
Kolmogorovian probability theory.

\subsection{Quantum Case}

The quantum case can be described in an analogous way. If
$\mathbf{A}$ represents the self adjoint operator of an observable
associated to a quantum particle, the proposition ``the value of
$\mathbf{A}$ lies in the interval $\Delta$" will define an event
represented by the projection operator
$\mathbf{P}_{\mathbf{A}}(\Delta)\in\mathcal{P}(\mathcal{H})$, i.e.,
the projection that the spectral measure of $\mathbf{A}$ assigns to
the Borel set $\Delta$. The probability assigned to the event
$\mathbf{P}_{\mathbf{A}}(\Delta)$, given that the system is prepared
in the state $\rho$, is computed using the Born's rule:
$p(\mathbf{P}_{\mathbf{A}}(\Delta))=\mbox{tr}(\rho\mathbf{P}_{\mathbf{A}}(\Delta))$.
Born's rule defines a measure on $\mathcal{P}(\mathcal{H})$ with
which it is possible to compute all probabilities and mean values
for all physical observables
\cite{DallaChiara-Giuntini-Greechie-Book,Birkhoff-vonNeumann-LogicOfQm}.
As an example, consider the energy of a quantum harmonic oscillator.
The proposition ``the energy of the oscillator equals
$\varepsilon_{i}$", corresponds to the projection operator
associated to the eigenspace of the eigenvalue $\varepsilon_{i}$.

It is well known that, due to Gleason's theorem \cite{Gleason1975},
a quantum state will be defined by a measure $s$ over the
orthomodular lattice of projection operators
$\mathcal{P}(\mathcal{H})$ as follows
\cite{RedeiSummersQuantumProbability}:

\begin{subequations}\label{e:nonkolmogorov}
\begin{equation}
s:\mathcal{P}(\mathcal{H})\rightarrow [0;1]
\end{equation}
\noindent such that:
\begin{equation}\label{e:Qprobability1}
s(\textbf{0})=0 \,\, (\textbf{0}\,\, \mbox{is the null subspace}).
\end{equation}
\begin{equation}\label{e:Qprobability2}
s(P^{\bot})=1-s(P),\end{equation} \noindent and, for a denumerable
and pairwise orthogonal family of projections ${P_{j}}$
\begin{equation}\label{e:Qprobability3}
s(\sum_{j}P_{j})=\sum_{j}s(P_{j}).
\end{equation}
\end{subequations}

\subsection{General Case}

\noindent Notice that despite their similarities, the difference
between \eqref{e:kolmogorovian} and \eqref{e:nonkolmogorov} is that
$\Sigma$ is replaced by $\mathcal{P}(\mathcal{H})$, and the other
conditions are the natural generalizations of the clasical event
structure to the non-Boolean setting. A general probabilistic
framework
---encompassing the Kolmogorovian and the quantal cases--- can be described
by the following equations

\begin{subequations}
\begin{equation}\label{e:GeneralizedProbability}
s:\mathcal{L}\rightarrow [0;1],
\end{equation}

\noindent ($\mathcal{L}$ standing for the lattice of all events)
such that:

\begin{equation}\label{e:Qprobability1}
s(\textbf{0})=0.
\end{equation}

\begin{equation}\label{e:Qprobability2}
s(E^{\bot})=1-s(E),\end{equation} \noindent and, for a denumerable
and pairwise orthogonal family of events $E_{j}$
\begin{equation}\label{e:Qprobability3}
s(\sum_{j}E_{j})=\sum_{j}s(E_{j}).
\end{equation}
\end{subequations}

\noindent where $\mathcal{L}$ is a general orthomodular lattice
(with $\mathcal{L}=\Sigma$ and
$\mathcal{L}=\mathcal{P}(\mathcal{H})$ for the Kolmogorovian and
quantum cases respectively). Eqns.
\eqref{e:GeneralizedProbability} define what is known as a
\emph{generalized probability theory}. Discussing the conditions
under which the measure $s$ in Eqns.
\eqref{e:GeneralizedProbability} is well defined (for very general
orthomodular lattices), lies outside  the scope of this paper; for
a detailed discussion see \cite{BeltramettiCassinelliBook},
Chapter $11$. It will suffice for us to notice that many examples
of interest in physics, including non-relativistic and
relativistic QM, and many examples of classical and quantum
statistical physics, can be described using orthomodular lattices
of projections arising from factors of Type I, II, and III , for
which measures such as those defined by Eqs.
\eqref{e:GeneralizedProbability} are well defined
\cite{RedeiQuantumLogicInAlgebraicApproach,RedeiSummersQuantumProbability}.

\emph{In the following Sections, we will develop a theoretical
framework which combines geometric probability theory, generalized
probability theory, and the Jayne's MaxEnt method.}

\section{A new set of axioms for physical
problems}\label{s:NewSetOfAxioms}

\subsection{Classical States As Invariant Measures}

Suppose that we are faced with the problem of determining the
particular probabilistic state $\mu$ of a classical system $S$. In
order to determine $\mu$, we must use the fact that it is a
probability measure over the event space $\mathcal{P}(\Gamma)$.
Thus, it will obey Eqns. \eqref{e:kolmogorovian}, which are
equivalent to the first two Axioms of Rotta (Eqs.
\eqref{a:RotaAxiom1} and \eqref{a:RotaAxiom2}) plus the
sigma-additivity condition \eqref{e:sigmaAdditivityKolomogorov}.
Imposing \textbf{Axiom} \ref{a:RotaAxiom3} entails that our system
would be in a state which possesses the symmetry of being invariant
under the whole group $\mathrm{E}_{0}$ of translations and rotations
of $\mathcal{P}(\Gamma)$. Call $\mathrm{E}$ to the group of all
possible Galilean transformations acting on the system (notice that
$\mathrm{E}_{0}\subseteq\mathrm{E}$). In the general case, the state
will not be invariant under all the elements of $\mathrm{E}_{0}$,
but will be invariant under a subgroup
$\mathrm{G}\subseteq\mathrm{E}$ (which could be just the identity
group, $\{\mathbf{1}\}$). For example, equilibrium states of a
system with cylindric symmetry will typically be invariant under
rotations and translations along $\hat{z}$ axis, but not for all
possible rotations and translations. We will use these observations
to generalize Rota's axioms.

Thus, a classical system will have probabilities obeying an
alteration of the Rota axioms. In it, i) $\mathrm{E}_{0}$ in
\textbf{Axiom} \eqref{a:RotaAxiom3} is replaced by a general
subgroup $\mathrm{G}\subseteq\mathrm{E}$, and ii) axiom
\ref{a:RotaAxiom4} is replaced by a series of conditions of the form

\begin{equation}\label{a:ConditionClassical1}
\langle f_{i}\rangle=r_{i},
\end{equation}

\noindent which represent the mean values of observables that are
available as empirical data. The group $\mathrm{H}$ and conditions
\eqref{a:ConditionClassical1} represent the a priori information
that we have regarding the system (notice that, to the traditional
prior information of the Jaynes's method expressed as mean values,
we are adding the possibility of symmetry constrains).

Thus, in order to determine the state $\mu$ of the system, we must
first solve the problem of determining the measures which satisfy
the usual probability axioms, plus i) the condition of being
invariant under the group $\mathrm{G}$ and ii) satisfying the
condition given by Eqn. \eqref{a:ConditionClassical1}. \emph{In this
way, the problem of  handling geometric probability can be
transformed into a physical one.}

\subsection{Quantum States As Invariant Measures}

Let us concentrate now on the quantum case before we turn to the
general setting. (Continuous) symmetry transformations in QM are
represented by the elements of the group of unitary operators
$\mathcal{U}$ \cite{VaradarajanBookI}. If we know in advance that
the state that we are looking for possesses a certain symmetry, this
condition will be represented by the invariance of the state under
the action of a subgroup $\mathcal{G}\subseteq\mathcal{U}$. Next, a
series of conditions on mean values of observables can be added.
These can be either mean values of operators or more general ones,
but which are insufficient on their own to fully determine the
state. These conditions can be cast in the form

\begin{equation}\label{e:ConditionsQuantum}
\langle \mathbf{A}_{i}\rangle=a_{i}
\end{equation}

A state will be represented by a \emph{measure} $s$ over the event
structure $\mathcal{P}(\mathcal{H})$. In other words, we are looking
for a measure $s$ which i) satisfies Eqns. \eqref{e:nonkolmogorov},
ii) that is \emph{invariant} under the action of the group
$\mathcal{G}$, and iii) satisfies Eqns. \eqref{e:ConditionsQuantum}.
Thus, in order to determine a quantum state compatible with the
prior knowledge about symmetries and mean values, we must determine
a measure such that the \textbf{Axioms} \eqref{a:RotaAxiom3} and
\eqref{a:RotaAxiom4} be adequately modified.

We see that, as in the classical case, the Rotta's problem can be
extended to the problem of determining the state of a physical
system, provided we generalize subsets of Euclidean space to the
lattice of projections in a Hilbert space, replace the
roto-translational group by the corresponding quantum one, and
replace the normalization condition by known mean values of a
given set of observables. These conditions restrict the possible
states to a subset of the space of quantum states. Following
Jaynes \cite{Jaynes-1957b} now, the least biased probability
distribution can be determined  by maximizing von Neumann's
entropy in this subset. It is nice that these observations are
susceptible of an even greater degree of generalization.

\subsection{Invariant Measures In Generalized
Theories}\label{s:GeneralizedAxioms}

Now we pass to  a systematic generalization of the above procedure
for quite arbitrary statistical theories, which will provide a new
ground for the MaxEnt principle. In this vein, we are led to
formulate the following set of axioms for a general physical
system, incorporating prior knowledge about symmetries and
conditions on expectation values (or even more general
conditions). The objective is to determine the unknown state $s$
of given system as an invariant measure obeying Eqs.
\eqref{e:nonkolmogorov}.

\noindent \textbf{Symmetries}: Knowledge about symmetries of the
physical system will be represented by the existence of a subgroup
$\mathfrak{G}$ of the group automorphisms of $\mathcal{L}$,
$Aut(\mathcal{L})$, such that for all $g\in\mathfrak{G}$, and for
all $E\in\mathcal{L}$,

\begin{equation}\label{e:GroupQprobability}
s(g\cdot E)=s(E).
\end{equation}

\noindent \textbf{Normalization condition}: There exists a set of
equations $\{e_i\}_I$ in the values $\{s(E_j)\}_J$,
\begin{equation}\label{e:ConditionsOnQprobability}
e_i(s(E_{1}),s(E_2),\ldots)=0,
\end{equation}
where $\{E_j\}_J\subseteq\mathcal{L}$ is some subset of events.

To summarize, we set down all the axioms that the unknown state
$\nu$
---now considered as a generalized invariant measure $\nu:\mathcal{L}\rightarrow [0;1]$ over an arbitrary
orthomodular lattice $\mathcal{L}$--- must satisfy:

\begin{axiom}[G1]\label{e:Generalized1}
$$\nu(\textbf{0})=0$$
\end{axiom}

\begin{axiom}[G2]\label{e:Generalized2}
$$\nu(E^{\bot})=1-s(E)$$,
\end{axiom}

\begin{axiom}[G3]\label{e:Generalized3}
\noindent For a denumerable and pairwise orthogonal family of events
$E_{j}$,
$$\nu(\sum_{j}E_{j})=\sum_{j}\nu(E_{j})$$.
\end{axiom}

\begin{axiom}[G4]\label{e:Generalized4}
\noindent For all $g\in\mathfrak{G}$
$$\nu(g\cdot E)=\nu(E)$$.
\end{axiom}

\begin{axiom}[G5]\label{e:Generalized5}
\noindent There exists a family of events $\{E_{j}\}$ which satisfy
the equations defined by functions $e_{i}$
$$e_i(\nu(E_{1}),\nu(E_2),\ldots,\nu{E_{m_{i}}})=0$$.
\end{axiom}
48625636 \noindent The above \textbf{Axioms} represent our
generalization of geometric probability to the noncommutative case.
\textbf{Axioms} \eqref{e:Generalized1}, and \eqref{e:Generalized2}
and \eqref{e:Generalized3} univocally determine a convex set
$\mathcal{S}$ (provided that $\nu$ be well defined, cf.
\cite{BeltramettiCassinelliBook}, Chapter \textbf{11}). It is
important to remark that the introduction of \textbf{Axiom}
\eqref{e:Generalized4} yields a smaller set
$\mathcal{S}_{\mathfrak{G}}\subseteq\mathcal{S}$ which is also
convex. The addition of \textbf{Axiom} \eqref{e:Generalized5}
determines a manifold $\mathcal{M}$, which, when intersected with
$\mathcal{S}_{\mathfrak{G}}$, will not necessarily yield a convex
set. However, it can be shown that if the constraints are mean
values imposed on observables, or more generally, on effects, the
set determined by $\mathcal{S}_{\mathfrak{G}}\cap\mathcal{M}$ will
be convex \cite{Holik-Plastino-QuantalEffects-MaxEnt}. Thus, the set
of states compatible with the prior knowledge about symmetries and
measured quantities will be the intersection
$\mathcal{S}_{\mathfrak{G}}\cap\mathcal{M}$.

Once this set is determined, Jaynes's entropic maximization process
singles out the less unbiased state which will rule the
probabilities of the system. In the following \textbf{Section}, we
discuss which entropic measures are to be used for this purpose.
Notice that if if $\mathcal{S}$ is compact, then
$\mathcal{S}_{\mathfrak{G}}$ and
$\mathcal{S}_{\mathfrak{G}}\cap\mathcal{M}$ will be also compact,
and we can ensure the existence of a solution for the maximization
procedure (provided that the entropic measure that we use be
continuous). Many physical examples comply with these assumptions
(for example, in non-relativistic quantum mechanics, the state space
is compact and the symmetry groups are locally compact).

\subsection{Entropies}

We wish  to define a meaningful notion of entropy for using it in
several frameworks, in the sense of being applicable to QM,
classical mechanics, and to general theories. Thus, we need an
appropriate notion of information measure, to be applied to general
statistical theories. One possibility is to use the so called
\emph{measurement entropy}, which reduces to Shannon's measure for
classical models and to von Neumann's in the quantum case
\cite{Barnum2010,HeinEntropyInQL1979}. Let $s$ be a state in a
generalized probability theory. Then, following Ref.
\cite{Barnum2010}, we define

\begin{equation}
H_{E}(\nu):=-\sum_{x\in E}\nu(x)\ln(\nu(x)),
\end{equation}

\begin{equation}
H(\nu):=\inf_{E\in \mathcal{L}}H_{E}(\nu).
\end{equation}
We show a comparison of the different cases in Table \ref{t:Table1}.

%
%

\begin{table}[h]
\begin{tabular}{l|l|l|l|}
\cline{2-4} & \,\,\,\,\,\,\,\,\,\textsc{Classical} &
\,\textsc{Quantum} & \,\,\,\,\,\textsc{General}
\\ \hline
\multicolumn{1}{|l|}{\textsc{Lattice}} &
\,\,\,\,\,\,\,\,\,\,\,\,\,\,\,$\mathcal{P}(\Gamma)$ &
\,\,\,\,\,\,\,\,\,$\mathcal{P}(\mathcal{H})$ &
\,\,\,\,\,\,\,\,\,\,\,\,\,$\mathcal{L}$
\\ \hline
\multicolumn{1}{|l|}{\textsc{Group}} &
\,\,\,\,\,\,\,\,\,\,\,\,\,\,\,$\mathrm{G}\subseteq \mathrm{E}$ &
\,\,\,\,\,\,\,$\mathcal{G}\subseteq \mathcal{U}$ &
\,\,\,$\mathfrak{G}\subseteq Aut(\mathcal{L})$
\\ \hline
\multicolumn{1}{|l|}{\textsc{Entropy}} & $-\sum_{i}p(i)\ln(p(i))$ &
$-\mbox{tr}\rho\ln(\rho)$ & $\inf_{E\in \mathcal{L}}H_{E}(\nu)$
\\ \hline
\end{tabular}
\caption{Table comparing the differences between the classical,
quantal, and general cases.} \label{t:Table1}
\end{table}

\subsection{Frame Functions And Group Actions}

Assume that a group $\mathfrak{G}$ is acting by automorphisms on a
lattice of events $\mathcal{L}$, $G\subseteq Aut(\mathcal{L})$
\cite{VaradarajanBookI,VaradarajanBookII}. Consider the convex set
$\mathcal{S}$ of Section \ref{s:GeneralizedAxioms}. \textbf{Axiom}
\eqref{e:Generalized4} states that invariant states are constant
along the orbits of the action,

$$s(g\cdot E)=s(E),\quad g\in \mathfrak{G},\, E\in \mathcal{L},$$

\noindent and an invariant state in $\mathcal{L}$ defines in a
canonical way a state in $\mathcal{L}/\mathfrak{G}$, where
$\mathcal{L}/\mathfrak{G}$ is the quotient lattice.

Assume now that the lattice $\mathcal{L}$ is atomic, where the set
of atoms is an $n$-dimensional compact manifold $\mathcal{A}$.
According to Gleason \cite{Gleason1975}, a state in $\mathcal{L}$ is
determined by a \emph{frame function} in $\mathcal{A}$, that is,
$$f:\mathcal{A}\rightarrow\mathbb{R},\quad \sum_{i=1}^r f(x_i)=1,$$
where $\{x_1,\ldots,x_r\}$ is a set in $\mathcal{L}$ such that
$x_i\bot x_j$ ($i\neq j$) and $x_1\vee\ldots\vee x_r=1$. Call
$\mathcal{F}$ to the set of frame functions. The full group of
automorphisms of the atomic lattice, $Aut(\mathcal{L})$, induces an
action in $\mathcal{A}$ and $\mathcal{F}$ is stable under this
action. If $f\in\mathcal{F}$ and $g\in Aut(\mathcal{L})$, then
$g\cdot f$ is also a frame function. Note that the continuous frame
functions $\mathcal{F}_{cont}\subseteq \mathcal{F}$ is a subset of
all the bounded continuous functions in $\mathcal{A}$,
$F_{cont}\subseteq L^{\infty}(\mathcal{A})$, and that the polynomial
frame functions are dense in $F_{cont}$.

The action of the group $G$ in $\mathcal{L}$ restricts itself to an
action on $\mathcal{A}$, and a frame function determines an
invariant state if and only if the frame function is invariant,
$g\cdot f=f$, for all $g\in \mathfrak{G}$. Thus, the invariant
states are characterized by the frame functions in
$\mathcal{A}/\mathfrak{G}$. Recall that the dimension of
$\mathcal{A}/\mathfrak{G}$ is equal to the dimension of
$\mathcal{A}$ minus the dimension of an orbit.

As  an example, consider an $(n+1)$-dimensional Hilbert space and
its lattice of subspaces, $\mathcal{L}$. The set of atoms (the rays
in the Hilbert space) is a projective space $\mathbb{P}^n$. It is a
compact variety of dimension $n$, $\mathcal{A}=\mathbb{P}^n$. The
full group of automorphisms of $\mathcal{L}$ is the Lie group
$\mathcal{U}$. In \cite{Gleason1975}, the fact that the set of frame
functions $\mathcal{F}$ is stable under $\mathcal{U}$ is used to
characterize frame functions as density matrices (positive
semi-definite self-adjoint operators of the trace class).

Consider now a group $\mathcal{G}\subseteq \mathcal{U}$, acting on
$\mathbb{P}^n$, and let us consider states invariant under the group
$\mathcal{G}$. Given that states are characterized by density
matrices, the invariant states are density matrices stable under
$\mathcal{G}$
$$\rho=g\cdot \rho,\quad\forall g\in \mathcal{G},$$
or equivalently, frame functions in $\mathbb{P}^n/\mathcal{G}$. Note
that we are reducing the dimension of the convex set of states and
the reduction will depend on the nature of the action of
$\mathcal{G}$.

\section{Covariance and symmetries}\label{s:CovarianceAndSymmetries}

A space time symmetry will have an action on the observables of the
system and on the state space. But this implies at the same time
that it will have an action on the associated operational logic. As
an example, consider the Galilei group in non-relativistic QM. Any
operator of the group acts on the variety of space time observables
(position, momentum) but at the same time there exists a
representation of this group in the set of unitary operators of
Hilbert space. Indeed, the content of Wigner's theorem asserts that
symmetry transformation preserving probabilities will have a
representation as a unitary or anti-unitary operator in Hilbert
space. This means that for each symmetry, say, a rotation, there
exists an automorphism acting on the logic of projection operators.

Thus, symmetries are usually generalized as follows \footnote{This
methodology can be traced back to \cite{LudwigBookI},
\cite{LudwigBookII}, \cite{VaradarajanBookI},
\cite{VaradarajanBookII} and \cite{MackeyFoundationsOfQM}.}. Suppose
that we have a group $\mathfrak{G}$ representing symmetries of a
physical system. Call $\mathcal{S}$ the set of all probability
measures. The elements of $\mathfrak{G}$ will also induce
transformations in $\mathcal{S}$ as convex automorphisms. As it is
well known \cite{VaradarajanBookI,VaradarajanBookII}, this group
will also have a representation in $Aut(\mathcal{L})$. Thus, for any
element $g\in\mathfrak{G}$, any event $E\in\mathcal{L}$ and any
$\nu\in\mathcal{S}$, a symmetry of the system will satisfy the
covariance condition

\begin{equation}\label{e:CovarianceCondition}
\nu(E)=\nu'(E'),
\end{equation}
\noindent where $E'=g\cdot E$ and $\nu'=g\cdot \nu$.

\noindent The above equation is  important  for two main reasons:

\begin{itemize}
\item It allows us to incorporate into our system the very important
notion of representation of groups, acting as convex automorphisms
on $\mathcal{S}$ and automorphisms of $\mathcal{L}$. The action of
these groups represents the actions of symmetry transformations
(including the spatiotemporal ones) and imposes conditions on the
geometry of $\mathcal{S}$ and observable algebras.
\item We will use this approach to define coherent states in the
general setting. First, because the introduction of symmetries
obeying the covariance condition \eqref{e:CovarianceCondition}
allows for the definition of a base state (as is the case for the
vacuum state of the electromagnetic field). Secondly, because the
group axiom allows us to pick up only those measures which satisfy
the condition of being coherent states.
\end{itemize}

\section{Coherent states}\label{s:CoherentStates}

Given the Heisenberg uncertainty relation in a state $\rho$

\begin{equation}\label{e:Uncertainty}
\Delta\mathbf{P}\Delta\mathbf{Q} \geq \frac{\hbar}{2}
\end{equation}

\noindent where for an operator $\mathbf{O}$,
$\Delta\mathbf{O}=\sqrt{\langle\mathbf{O}^{2}\rangle-\langle\mathbf{O}\rangle^{2}}$,
coherent states
\cite{Glauber1963a,Glauber1963b,Zhang-Feng-CoherentStatesReview} are
defined as those which saturate \eqref{e:Uncertainty} with equal
mean values, i.e.:

\begin{subequations}\label{e:UncertaintySaturated}
\begin{equation}
\Delta\mathbf{P}\Delta\mathbf{Q}=\frac{\hbar}{2}
\end{equation}
\begin{equation}\label{e:EqualMean}
\Delta\mathbf{P}=\sqrt{\frac{\hbar}{2}}=\Delta\mathbf{Q}
\end{equation}
\end{subequations}

\noindent Thus, we can easily incorporate such states into our
conceptual framework by replacing \eqref{e:Generalized4} by Eqs.
\eqref{e:UncertaintySaturated}. Note that Eq. \eqref{e:EqualMean}
produces a real algebraic variety $\mathcal{M}$ in the real vector
space of Hermitian operators (it is given by the zero locus of the
two polynomial equations of degree two,
$\Delta\mathbf{P}=\sqrt{\frac{\hbar}{2}}$ and
$\Delta\mathbf{Q}=\sqrt{\frac{\hbar}{2}}$).

In arbitrary dimension ($n\leq\infty$) the states satisfying Eq.
\eqref{e:EqualMean} are given by the intersection of two quadrics.
Recall that any quadric can be parameterized an thus the
intersection $\mathcal{C}\cap\mathcal{M}$ can be computed in finite
dimensions. If the convex set $\mathcal{C}$ is a compact set, then
the intersection $\mathcal{C}\cap\mathcal{M}$ is also compact.

We can also define coherent states using group theory. This has the
advantage of being easily applicable to general statistical theories
\footnote{It is important to remark here that both  the definition
of coherent states that uses Eqs. \eqref{e:UncertaintySaturated} and
the group theoretical one are equivalent for the case of the
electromagnetic field, but will not be equivalent in general (as is
the case for finite dimensional Hilbert spaces)
\cite{Zhang-Feng-CoherentStatesReview}. Thus, it is not expected
that these definitions will be equivalent in arbitrary statistical
theories neither.}. While the choice of a reference state $s_{0}$
is, in principle, arbitrary \cite{Zhang-Feng-CoherentStatesReview},
the use of physical symmetries could be useful for its
determination. These will be represented by a group action
$\mathfrak{G}$ which, as mentioned above, induces actions in
$\mathcal{L}$ and $\mathcal{S}$. This procedure singles out the
correct reference state $s_{0}$
\cite{Zhang-Feng-CoherentStatesReview} by using the generalization
of geometric probability described in previous Sections. Once
$s_{0}$ is specified, we invoke the action of a given dynamical
group $G$, determine its maximum stability subgroup $H$
\cite{Zhang-Feng-CoherentStatesReview}, and construct the set of all
coherent states $\mathcal{S}_{G}\subseteq\mathcal{S}$ as follows

\begin{equation}
s_{g} := g\cdot s_0,
\end{equation}

\noindent where $g$ ranges over all the elements of
$G/H$\cite{Zhang-Feng-CoherentStatesReview}.

\section{Bell inequalities, no-signal polytope and local
polytope}\label{s:BellInequalities}

Immense interest generates in the study of  correlations in QM.
For two separate observers, $A$ and $B$, both of them having
available two observables $\{a_{0},a_{1}\}$ and $\{b_{0},b_{1}\}$,
with two possible outcomes for each, the correlations will be
governed by probability distributions of the form
$P(a_{i},b_{j}|x,y)$. It can be shown that the following
inequalities can be violated by QM

\begin{equation}\label{e:CHSH}
S=|\langle a_{0}b_{0}\rangle+\langle a_{1}b_{0}\rangle+\langle
a_{0}b_{1}\rangle-\langle a_{1}b_{1}\rangle|\leq 2,
\end{equation}

\noindent These are known as the Clauser-Horne-Shimony-Holt (CHSH)
inequalities \cite{Bruner-2014-ReviewBell,PopescuNature-2014}. The
no-signal polytope, formed by all possible correlations respecting
the no-signal condition of special relativity, is defined by the
following conditions \cite{PopescuNature-2014}

\begin{subequations}\label{e:NoSignal}
\begin{equation}
\sum_{j} P(a_{i},b_{j}|x,y) = \sum_{j} P(a_{i},b_{j}|x,y')\,\,\,
\forall y,y'
\end{equation}
\begin{equation}
\sum_{i} P(a_{i},b_{j}|x,y)=\sum_{i}P(a_{i},b_{j}|x',y)\,\,\,
\forall y,y'
\end{equation}
\end{subequations}

\noindent Quantum correlations can violate the CSHS inequalities,
but at the same time, they respect the no-signal condition (the
distributions $P(a,b|x,y)$ lie inside the no-signal polytope). One
may ask which is the characteristic trait of quantum mechanics
that distinguishes it from  general statistical theories which are
also no-signal, but do not produce the correlations predicted by
QM \cite{PopescuNature-2014}. This issue can be studied within our
theoretical framework by setting conditions \eqref{e:CHSH} and
\eqref{e:NoSignal} as axioms in the event space. By replacing
condition \eqref{e:Generalized5} by \eqref{e:CHSH}, we obtain the
local polytope, and by replacing it by \eqref{e:NoSignal}, we
obtain the no-signal polytope. The reformulation of these
geometrical objects within  our framework could permit the study
of the action of suitable groups of space-time symmetries (by
introducing these groups through Axiom \eqref{e:Generalized4}).




\section{Conclusions}\label{s:Conclusions}

It is important to remark that a systematic presentation of the
Jaynes's method, as we have done here, has  not yet be advanced in
the literature, as far as we know. We summarize our conclusions as
follows:

\begin{itemize}
\item The use of a formulation based on the traditional axiomatic method of measure theory
allows for a rigorous approach to MaxEnt.
\item We have shown that many cases fall within the axiomatic
framework presented here (coherent states, no-signal polytopes,
local polytopes). When our group symmetry is reduced to the identity
and the constraints are expressed as mean values, our method reduces
to previous generalizations of the Jaynes's methodology
\cite{Holik-Plastino-QuantalEffects-MaxEnt,HeinEntropyInQL1979}.

\item When $\mathcal{L}=\mathcal{L}_{v\mathcal{N}}$ or
$\mathcal{L}=\mathcal{P}(\Gamma)$, and the constraints are
expressed as mean values, our method reduces to the pioneer
Jaynes's one for the quantum and classical cases, respectively.

\item Our rigorous formulation allows us to establish
precise conditions for the existence of solutions to the MaxEnt
problem for very general constraints (including group theory,
non-linear conditions on the mean values of observables, and
inequalities as well).

\item At the same time, we provide an intrinsic
geometric characterization for the different mathematical objects
defined within our theoretical framework (quadrics for coherent
states, a convex set for the local polytope, etc.). Notice that this
may be of help in studying the geometrical properties of the
non-signal and local polytopes for the most general case (a
continuous range of observables with possibly continuous spectra. In
QM, in infinite dimensional Hilbert spaces). Our formulation may
help to extrapolate, in the future, results from Geometric
Probability Theory to physics.

\item By reformulating the problem in terms of the determination of
invariant measures, we provide a natural framework for the
introduction of group theory. We have explicitly shown that the
introduction of groups reduces the dimensionality of the
mathematical variety in which the maximization process takes place.
Thus, our proposal may be useful to economize computational
resources. Our axioms allow one to incorporate into the Jaynes's
framework the symmetries of the physical system under study. For
example, one could insert a group representing a spacial symmetry of
a system. This method yields a powerful resource for deriving laws
of physics out of general physical principles.

\item The facts that i) probability theory is a well established
theory and ii) explicit solutions to our problem can be found (as in
the examples studied in this work) show that our mathematical
problem is meaningful. This fact constitutes a clear improvement on
the MaxEnt method, giving a step forward into its axiomatization.

\end{itemize}

\begin{acknowledgements}

This work has been supported by CONICET.

\end{acknowledgements}

\bibliographystyle{apsrev4-1}
\bibliography{CarloRotaSubmitArXiv}

\end{document}